\documentclass[fleqn,usenatbib,useAMS]{mnras}

\usepackage{graphicx}	
\usepackage{amsmath}	
\usepackage{amssymb}	
\usepackage{multicol}        
\usepackage{bm}		
\usepackage{pdflscape}	
\usepackage[encapsulated]{CJK}
\usepackage{ucs}
\usepackage[utf8x]{inputenc}
\usepackage{multirow}
\usepackage{booktabs,caption}
\usepackage[flushleft]{threeparttable}
\usepackage[dvipsnames]{xcolor}
\usepackage{adjustbox}
\usepackage{soul}

\newcommand{\object}[1]{#1}

\newcommand{\kms}{\,km\,s$^{-1}$} 

\newcommand{\NVred}{N\,{\sc v}\,$\lambda 4945$}
\newcommand{\NVblue}{N\,{\sc v} $\lambda \lambda 4604, 4620$}

\usepackage[T1]{fontenc}
\usepackage{ae,aecompl}

\usepackage{newtxtext,newtxmath}

\title[On the variability in WR\,7]
{Multiple variability time-scales of the early nitrogen-rich Wolf-Rayet star WR\,7}
\author[Toal\'{a} et al.]{J.~A.\,Toal\'{a}$^{1}$\thanks{E-mail:\,j.toala@irya.unam.mx}, D.\,M. Bowman$^{2}$,  T.\,Van Reeth$^{2}$, H.\,Todt$^{3}$, K.\,Dsilva$^{2}$, T.\,Shenar$^{2}$, G.\,Koenigsberger$^{4}$,   
\newauthor{S.\,Estrada-Dorado$^{1}$, L.~M.\,Oskinova$^{3}$ and W.-R.\,Hamann$^{3}$}\\
$^{1}$Instituto de Radioastronom\'{i}a y Astrof\'{i}sica, UNAM Campus Morelia, Apartado postal 3-72, 58090 Morelia, Michoac\'{a}n, Mexico\\
$^{2}$Institute of Astronomy, KU Leuven, Celestijnlaan 200D, 3001 Leuven, Belgium\\
$^{3}$Institute for Physics and Astronomy, Universit\"{a}t Potsdam, Karl-Liebknecht-Str. 24/25, D-14476 Potsdam, Germany\\
$^{4}$Instituto de Ciencias F\'{i}sicas, Universidad Nacional Aut\'{o}noma de M\'{e}xico, Ave. Universidad s/n, Chamilpa, Cuernavaca, Mexico\\
}

\pubyear{2022}

\begin{document}
\label{firstpage}
\pagerange{\pageref{firstpage}--\pageref{lastpage}}
\maketitle

\begin{abstract}
\noindent 
We present the analysis of the optical variability of the early,
nitrogen-rich Wolf-Rayet (WR) star WR\,7. The analysis of multi-sector
{\it Transiting Exoplanet Survey Satellite} ({\it TESS}) light curves
and high-resolution spectroscopic observations confirm multi-periodic
variability that is modulated on time-scales of years. We detect a
dominant period of $2.6433 \pm 0.0005$\,d in the {\itshape TESS}
sectors 33 and 34 light curves in addition to the previously reported
high-frequency features from sector 7. We discuss the plausible
mechanisms that may be responsible for such variability in WR\,7,
including pulsations, binarity, co-rotating interacting regions (CIRs)
and clumpy winds. Given the lack of strong evidence for the presence
of a stellar or compact companion, we suggest that WR\,7 may pulsate
in quasi-coherent modes in addition to wind variability likely caused
by CIRs on top of stochastic low-frequency variability. WR\,7 is
certainly a worthy target for future monitoring in both spectroscopy
and photometry to sample both the short ($\lesssim 1$~d) and long
($\gtrsim 1000$~d) variability time scales.
\end{abstract}

\begin{keywords}
stars: evolution --- stars: atmospheres --- stars: winds, outflows --- stars: Wolf-Rayet --- stars: individual: WR\,7 
\end{keywords}




\section{INTRODUCTION}

\label{sec:intro}    

Wolf-Rayet (WR) stars represent one of the most advanced stages of
massive stellar evolution. They are H-depleted stars characterised by
the presence of broad emission lines from highly ionised species of
He, C, N, and O in their spectra \citep[see,
  e.g.,][]{Crowther2007}. WR stars are also distinguished by fast and
strong winds \citep[mass-loss rates
  $\dot{M}\approx$10$^{-5}$~M$_\odot$~yr$^{-1}$, terminal wind
  velocities $\varv_\infty\simeq$2000--3000~km~s$^{-1}$; e.g.,][and
  references therein]{Hamann2006,Hamann2019} which renders them the
hot stars with the most powerful winds.

The current formation scenario of WR stars is twofold. It has been
proposed that single massive stars with initial mass
$M_\mathrm{i}\geq$~20~M$_\odot$ might lose up to half their initial
masses via slow and dense winds when evolving through a red supergiant
(RSG) or eruptive ejections through a luminous blue variable (LBV)
phase \citep{Weis2001,Humphreys2010}. It is argued that this process
is able to peel off the outer H-rich layers, leaving behind a
classical WR star \citep{Conti1975}. According to the second scenario,
first proposed by \citet{Paczynski1967}, mass transfer between binary
components \citep{Packet1981,deMink2013} could cause stripping of the
H-rich envelope, a scenario which has recently gained impetus when it
was shown that the majority of massive stars evolve in binary systems
\citep[e.g.,][]{Vanbeveren1998,Foellmi2003,Sana2012,Sota2014}.

The binary evolutionary channels have interesting implications, for
example, extending the life times of the mass-donors and reducing the
initial mass necessary to form a WR star. Nevertheless, the dominance
of one scenario over the other has not been demonstrated
\citep[e.g.,][]{Neugent2014, Shenar2016, Shenar2020}, and a third
scenario in which a high internal mixing efficiency eventually leads
to a He-rich, non-stripped star cannot be excluded
\citep{Brott2011,Hainich2015}.

As the progenitors of neutron stars (NSs) and black holes (BHs), WR
stars in binaries offer indispensable laboratories to study the
evolution and formation of high-mass X-ray binaries (HMXBs) and BH
merger events \citep{Abbott2019}. In this context, it is important to
investigate and identify potential WR binaries with compact-object
companions.

The first such proposed object was WR\,6 (a.k.a. HD\,50896 or
EZ\,CMa). \citet{Firmani1979,Firmani1980} attributed the 3.7~d
periodic variations in the emission line profiles to the interaction
of the WR wind with a NS companion in an eccentric orbit. The
variations were subsequently attributed to other phenomena, including
two oppositely moving outflows \citep{Matthews1991}, an unspecified
mechanism due to rotation \citep{Drissen1989}, periodic instabilities
at the base of the wind \citep{Flores2007}, co-rotating interaction
regions \citep[CIRs;][]{Ignace2013} and the interaction of the WR wind
with a non-compact companion \citep{Koenigsberger2020}. Despite nearly
six decades of observational campaigns, there is as yet no theoretical
radiative transfer study that is capable of reproducing the line
profile variability that gave rise to these thus far speculative
scenarios.

Currently, only one candidate HMXB hosting a WR star and a BH was
identified in the Galaxy \citep[Cyg\,X-3;][]{Kerkwijk1992}, and a few
more in other galaxies \citep{Esposito2015}. The rarity of WR binaries
with compact object became known as the problem of "missing WR X-ray
binaries" \citep[e.g.,][]{VanDenHeuvel2017, Vanbeveren2020}, though it
has recently been argued that the majority of WR+compact-object
binaries are not expected to be X-ray bright \citep{Sen2021}, a
criticism that had been raised for WR\,6, for example. Thus, it is
therefore crucial to investigate whether other WR stars may be
candidates for harbouring a compact object companion.

Variability has been used historically as a tool to search for and
study the presence of companions
\citep[][]{Cherepashchuk1973,Moffat1986,Koenigsberger2020} and has
been further used to probe the wind properties in WR stars. As a
consequence of the line-driven instabilities
\citep{CAK1975,Owocki1988}, winds from WR stars are not expected to be
uniform. Inhomogeneities in their winds are expected to produce
time-dependent fluctuations in the light curves, a type of variability
also attributed at times to CIRs
\citep[see][]{StLouis2009,Cranmer1996}. Such structures are also
thought to be responsible for linear polarization
\citep{Robert1992,Fullard2020}. Other mechanisms that have been
proposed to cause wind variability are pulsations \citep[see][and
  references therein]{Townsend2007,Grassitelli2016}. In fact, the
connection between pulsations and outflowing wind structures has been
demonstrated for at least one well-studied object, HD\,64760
\citep{Kaufer2006}.

Among WR stars, those classified as part of the WN8 sequence are the
most variable \citep{Antokhin1995,Moffat1986,Marchenko1998}. Their low
temperatures produce a (sub)surface convection zone due to the iron
opacity bump \citep{Nugis2002,Cantiello2009}, which may have
consequences for their surface variability.  On the other hand, early
WN stars have stronger stellar winds and lower luminosities which make
them less affected by the iron opacity bump \citep{Ro2019}. As a
consequence, one could expect that such stars are less prone to
variability. However, variability has been also reported for early WN
stars in the past decades
\citep[e.g.,][]{Chene2010,Chene_etal_2011,Marchenko1998,Lamontagne1983,Koenigsberger1980,LenoirCraig2022}. Recently,
\citet{Naze2021} showed that there is no apparent dichotomy in the
variability between early and late WN stars.

In this paper we present the analysis of the optical variability from
the WN4 star WR\,7. Thus far, there are no suggestions of a companion
reported in the literature. Radio observations searching for
non-thermal emission as signatures of binary companions resulted only
in upper limits for WR\,7 \citep{Raguzova2000}. Furthermore, there are
no available UV or optical line variability studies of WR\,7 that show
the presence of CIRs. Recently, \citet{Naze2021} presented the
analysis of {\it Transit Exoplanet Survey Satellite} \citep[{\itshape
    TESS};][]{Ricker2015} observations of a sample of WR and LBV
stars, including WR\,7. They analysed Sector 7 {\it TESS} observations
of WR\,7 and concluded that it exhibits coherent variability with a
dominant period of 3.86~d$^{-1}$ with some harmonic
frequencies. Although they seem to suggest pulsations as the main
physical mechanisms to explain such periodicity, \citet{Naze2021}
conclude that new pulsating models are needed to enhance our
understanding of the observations.

In this paper we use {\it TESS} observations of WR\,7 to corroborate
multi-periodic variability of the order of days and hours. We
complement these TESS data with high-resolution time-series
spectroscopic observations to further study the variability of
WR\,7. In Section~\ref{sec:obs} we describe our observations and
results are presented in Section~\ref{sec:data_analysis}. We discuss
the possible origins of the variability in
Section~\ref{sec:discussion}. Finally, a summary is presented in
Section~\ref{sec:conclusions}.

\section{Observations}
\label{sec:obs}

The {\itshape TESS} satellite offers exciting and new opportunities to
study the photometric variability of many massive stars to an
unprecedented level of precision \citep{Bowman2020}. In an initial
search of variability in WR stars observed by {\it TESS}, we
corroborated WR\,7 as showing atypical properties. Specifically, its
light curve and corresponding amplitude spectrum contains both
long-period and significant short-period variability. The latter
caught our attention and motivated us to seek out plausible mechanisms
that may be responsible. {\it TESS} observed WR\,7 during sector 7
(between 2019 Jan 8 and 2019 Feb 1) in the full-frame images (FFIs)
with a 30-min cadence and during sectors 33 and 34 (between 2020 Dec
18 and 2021 Feb 8) both with a 2-min cadence and in the FFI with a
10-min cadence. We retrieved both 40$\times$40 pixel cutouts from the
FFIs with {\scshape Astrocut} \citep{Brasseur2019}, and short-cadence
target pixel files (TPFs) from the Mikulski Archive for Space
Telescopes (MAST)\footnote{\url{https://archive.stsci.edu/}}.

\begin{figure*}
\centering
\includegraphics[width=0.99\textwidth]{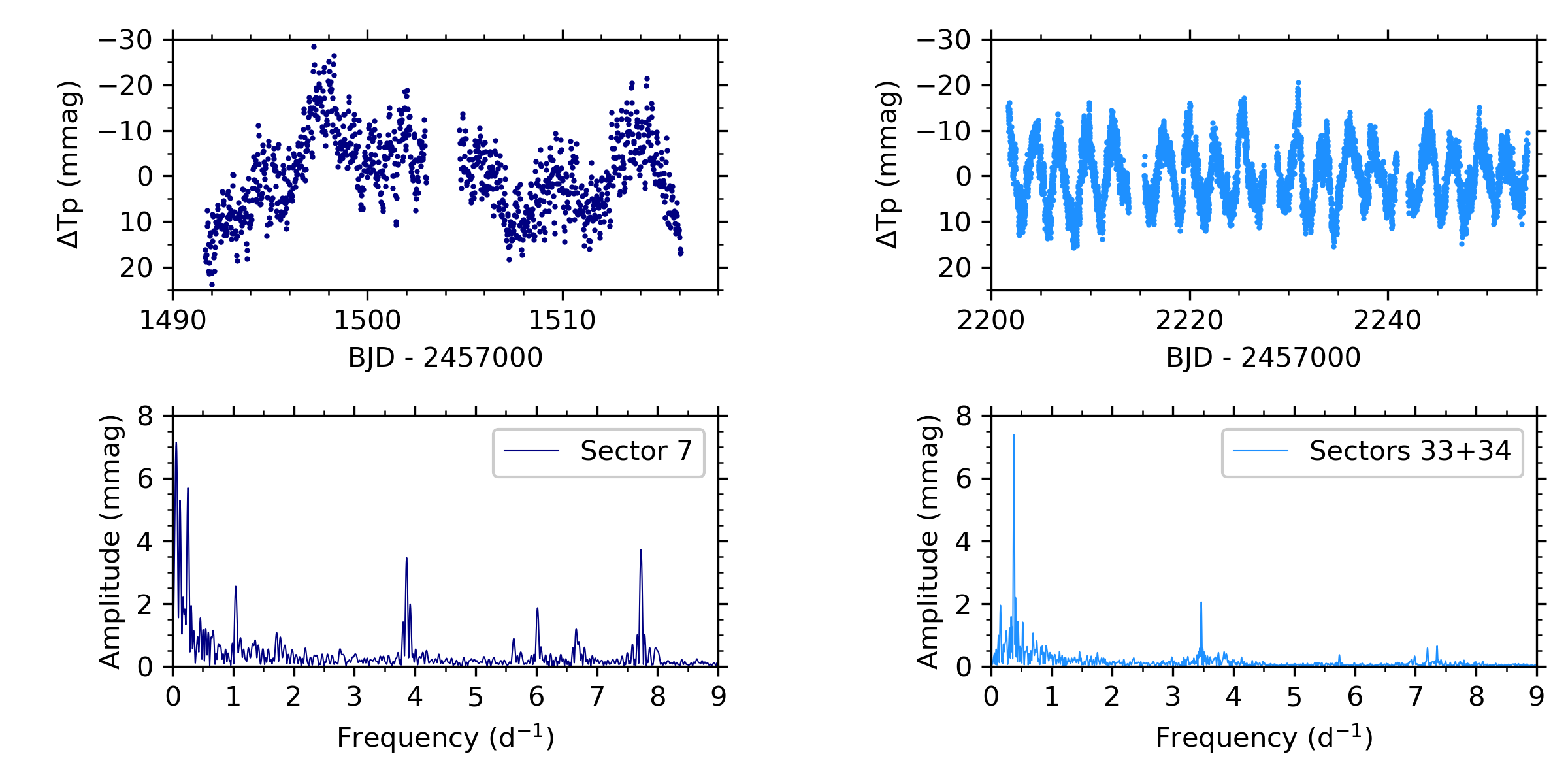}
\caption{Background-subtracted {\it TESS} light curves of WR\,7 (top
  panels) and their corresponding amplitude spectra (bottom panels) of
  our FFI data obtained in sector 7 (left) and sectors 33 and 34
  combined (right). Long-term modulation is clearly detected in both
  the long- and short-period variability in WR\,7.}
\label{fig:FT_grid}
\end{figure*}

We found that because the short-cadence TPFs are much smaller than our
own FFI pixel cutouts, their background flux estimates are
non-negligibly contaminated by the nearby eclipsing binary
OU~CMa. Hence, we focus ourselves to the FFI data in this work and
conclude that the light curves obtained from the SPOC pipeline applied
to the 2-min TPF data are inferior. We estimated the background flux
by taking the median observed flux per frame, excluding the pixels
that predominantly contain stellar flux, and subtracted it from the
measured flux values. The light curves were then obtained with
custom-created aperture masks, normalised by dividing through the
median flux and converted to units of mmag. Our FFI-extracted light
curves from sectors 7, 33 and 34, together with the corresponding
amplitude spectra, are shown in Fig.~\ref{fig:FT_grid}.

In addition to {\itshape TESS} data, we retrieved archival and
recently acquired high-resolution spectroscopy for \object{WR\,7}. The
first set of spectra were secured with the Ultraviolet and Visual
Echelle Spectrograph (UVES) instrument mounted on the UT2 of the Very
Large Telescope (VLT). A total of 14 spectra were retrieved in reduced
form from the European Southern Observatory (ESO) archive (program ID
080.D-0137, PI: Foellmi).  The spectra were acquired back-to-back in
the two nights of 2008 Jan 26--27 and have a nominal resolving power
of $R \approx 30\,000$.  We complemented these archival spectra with
seven spectra acquired with the HERMES spectrograph \citep{Raskin2011}
mounted on the 1.2~m Mercator telescope in La Palma, Spain. These
spectra have a nominal resolving power of $R \approx 80\,000$. The
spectra were acquired in the framework of a long-term spectroscopic
monitoring of Galactic WR stars \citep[see][for details]{Dsilva2020}.

\section{Data Analysis}
\label{sec:data_analysis}

\subsection{{\itshape TESS} photometry}\label{subsec:TessPhot}

\begin{figure*}
\centering
\includegraphics[width=\linewidth]{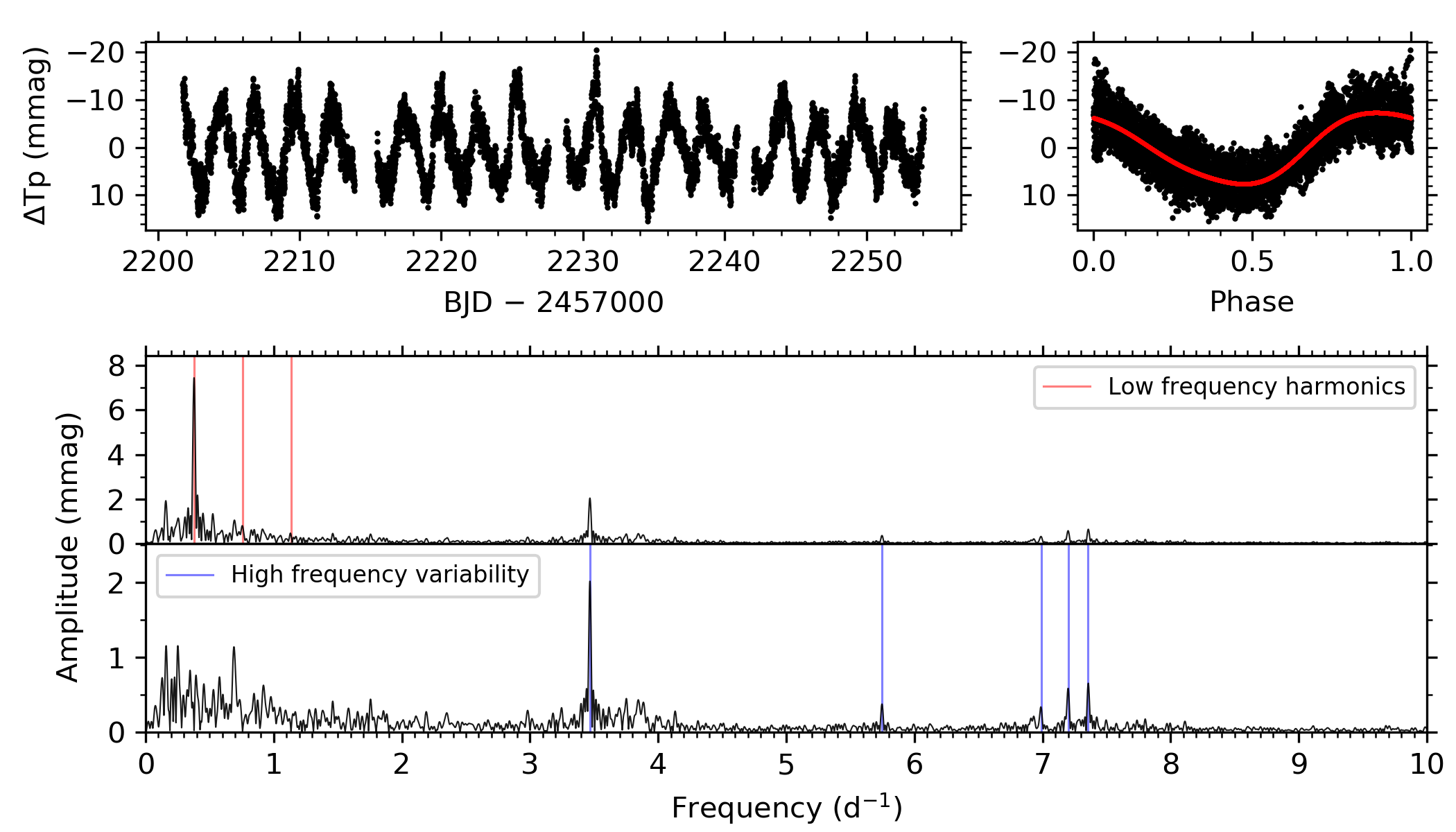}
\caption{Detrended 10-min FFI {\itshape TESS} light curve of WR\,7
  from sectors 33 and 34 (upper left panel), the phase-folded light
  curve (upper right panel) for the dominant period of 2.6433\,d, the
  amplitude spectrum (middle panel) and residual amplitude spectrum
  (bottom panel) after subtracting the dominant frequency and two
  significant harmonics, which are shown as vertical red
  lines. Vertical blue lines represent significant short period
  variability (cf. Table~\ref{tab:freq list}).}
\label{fig:TESS FT}
\end{figure*}

To assess the multiperiodic variability in the {\itshape TESS} light
curves of WR\,7, we used Fourier analysis to extract significant
frequencies, amplitudes, phases and their respective uncertainties. As
evidenced in Fig.~\ref{fig:FT_grid}, it became immediately apparent
that the variability seen in WR\,7 was different in sector 7 than in
sectors 33 and 34, which implies a long term modulation to the
variability given that these {\itshape TESS} sectors span almost
2~yr. Also, the {\itshape TESS} light curve from sector 7 is more
noisy, which may be astrophysical given that WR variability is known
to be dominated by stochastic signals associated with the clumpy
aspherical winds in combination with stellar pulsations as suggested
by \citet[][]{Naze2021}. We employed the detrending methodology
developed by \citet{Bowman2018} to the combined {\itshape TESS}
sectors 33 and 34 light curve, since we deem it of higher quality for
subsequent frequency analysis, to remove any remaining small
instrumental trends. The detrended FFI light curve from sectors 33 and
34 is shown in the top-left panel of Fig.~\ref{fig:TESS FT}.

\begin{table}
\begin{center}
\caption{Frequencies, amplitudes and phases of the significant
  (i.e. SNR $\geq 4$) signals in the {\itshape TESS} photometry of
  WR\,7. Because of the long-term amplitude and frequency modulation
  in WR\,7, the sector 7 and sectors 33 and 34 data are analysed
  separately. $1\sigma$ uncertainties are calculated from the
  multi-frequency non-linear least-squares fit, using the phase
  zero-points of BJD~2458505.0 and BJD~2459225.0, respectively.}
\begin{tabular}{rrr}
\hline
\multicolumn{1}{c}{Frequency} & \multicolumn{1}{c}{Amplitude} & \multicolumn{1}{c}{Phase} \\
\multicolumn{1}{c}{(d$^{-1}$)} & \multicolumn{1}{c}{(mmag)} & \multicolumn{1}{c}{(rad)} \\
\hline
\multicolumn{3}{l}{Sector 7} \\
$1.04378 \pm 0.00283$   &   $2.54 \pm 0.33$   &   $-2.39 \pm 0.13$    \\
$3.86187 \pm 0.00265$   &   $3.17 \pm 0.33$   &   $2.73 \pm 0.11$     \\
$3.92136 \pm 0.00734$   &   $1.14 \pm 0.33$   &   $-0.61 \pm 0.31$    \\
$5.62808 \pm 0.00923$   &   $0.78 \pm 0.33$   &   $1.18 \pm 0.43$     \\
$6.02026 \pm 0.00396$   &   $1.83 \pm 0.33$   &   $1.46 \pm 0.18$     \\
$6.65874 \pm 0.00628$   &   $1.15 \pm 0.33$   &   $3.03 \pm 0.29$     \\
$7.72781 \pm 0.00198$   &   $3.68 \pm 0.33$   &   $3.06 \pm 0.09$     \\
$7.98967 \pm 0.01404$   &   $0.52 \pm 0.33$   &   $2.03 \pm 0.65$     \\

\hline
\multicolumn{3}{l}{Sectors 33 and 34} \\
$0.37832 \pm 0.00007$   &   $7.45 \pm 0.05$   &   $2.03 \pm 0.01$     \\
$0.75424 \pm 0.00070$   &   $0.76 \pm 0.05$   &   $2.68 \pm 0.07$     \\
$1.12990 \pm 0.00136$   &   $0.39 \pm 0.05$   &   $-2.85 \pm 0.13$    \\
$3.46785 \pm 0.00026$   &   $2.02 \pm 0.05$   &   $-0.92 \pm 0.03$    \\
$5.74754 \pm 0.00144$   &   $0.37 \pm 0.05$   &   $2.56 \pm 0.14$     \\
$6.99000 \pm 0.00190$   &   $0.28 \pm 0.05$   &   $0.98 \pm 0.19$     \\
$7.20050 \pm 0.00087$   &   $0.62 \pm 0.05$   &   $0.81 \pm 0.09$     \\
$7.35643 \pm 0.00079$   &   $0.68 \pm 0.05$   &   $-2.15 \pm 0.08$    \\

\hline
\end{tabular}
\label{tab:freq list}
\end{center}
\end{table}  

To determine the significant periodicities and their amplitudes in the
{\itshape TESS} data, we fit sinusoids to the light curve using the
{\sc Period04} software \citep{Lenz2005}, opting to treat the sector 7
light curve and the combined sectors 33 and 34 light curve
separately. A significant frequency is defined as having an amplitude
signal-to-noise ratio (SNR) larger than four following the empirical
results of \citet{Breger1993}. The noise is estimated using a local
window of size 1~d$^{-1}$ centred in the amplitude spectrum at the
location of where a frequency has been removed during iterative
pre-whitening. At the end of iterative pre-whitening and all
significant (SNR~$\geq 4$) frequencies have been extracted, we
optimise their amplitudes, frequencies and phases using a cosinusoid
function and a non-linear least-squares fit to the light curve (see,
e.g., \citealt{Bowman2021}). In total we find eight significant
frequencies in the sector 7 light curve and a different eight
significant frequencies in the combined sectors 33 and 34 light curve,
which range between approximately 0.3 and 8.0~d$^{-1}$, and have
amplitudes up to 7.45~mmag. A complete list of significant
frequencies, including their optimised amplitudes, phases and
corresponding uncertainties from the sector 7 and combined light curve
from sectors 33 and 34, is given in Table~\ref{tab:freq list}.

Similarly to what is reported in \citet{Naze2021} for sector 7, we
found that the highest amplitude feature corresponds to a frequency of
$7.728 \pm 0.002$~d$^{-1}$, but this is very likely a harmonic of the
$3.862 \pm 0.003$~d$^{-1}$ frequency. That is, a dominant period of
$0.25894 \pm 0.0002$~d ($\approx$6.22~h). However, the longer time
base, larger number of data points, and improved noise statistics of
the combined sectors 33 and 34 light curve provides better precision
on the resultant frequencies compared to those of sector 7. In the
combined sectors 33 and 34 light curve, the dominant variability has a
frequency of $0.37832 \pm 0.00007$~d$^{-1}$, that is, a period of
$2.6433 \pm 0.0005$~d. Other significant frequencies include harmonics
of this period (see Table~\ref{tab:freq list}). The phase-folded light
curve from sectors 33 and 34 based on this dominant period and its two
significant harmonics is shown in the top-right panel of
Fig.~\ref{fig:TESS FT}. In the bottom panel, the original and residual
amplitude spectra (i.e. after subtracting the dominant period and its
two significant harmonics) are shown. The presence of harmonics
indicate the non-sinusoidal nature of the long-period variability, and
suggest rotation modulation or binarity as possible causes. The
vertical blue lines in the bottom panel of Fig.~\ref{fig:TESS FT}
denote the location of the significant high frequency variability in
the amplitude spectrum.

In general, we detect high-amplitude, high-frequency variability with
frequencies between 3--10~d$^{-1}$ that appear in both {\it TESS}
sector observations of WR\,7 which were previously inferred to be
associated with stellar pulsations \citep{Naze2021}.  Nevertheless,
the amplitudes and frequencies between sector 7 and those of sectors
33 and 34 light curves are significantly different (see
Table~\ref{tab:freq list} and Fig.~\ref{fig:FT_grid}). Not only do the
amplitude and frequencies of the short-period variability change
significantly, the dominant period of the 33+34 sector (2.64~d) is not
detectable in the sector 7 data. This clearly demonstrates that WR\,7
has multiple variability time-scales of order days and hours,
respectively, which are both modulated on the time scale of years
given the gap of $\sim$2~yr between the sectors 7 and the sectors 33
and 34 light curves. Such an apparent incoherence and clear modulation
of both long- and short-period variability in a WR is intriguing and
points to an astrophysical explanation. We note that epoch-dependent
measurements have also been reported by \citet{LenoirCraig2022} for
other galactic WR stars. Hence, such variability is difficult to
reconcile as coherent pulsations given its long-term modulation of
order months-to-years. On the other hand, coherent pulsations in the
photosphere of a WR star would be heavily obscured and strongly
modulated by the clumpy wind.

Moreover, we investigated if the variability seen in the {\itshape
  TESS} data of WR\,7 could be caused by contamination, and confirmed
that the observed variability is intrinsic to WR\,7. Aside from
small-scale variations depending on the chosen aperture mask, the
level of flux contamination in the final light curve is only
$\sim$5~per~cent. In addition, the observed amplitude of the
short-period variability is maximal at the location of WR\,7 on the
CCD and correlates with the flux contribution of WR\,7 within a given
pixel of the CCD, providing strong support for the variability
originating in WR\,7. This is assuming, however, that WR\,7 does not
have an as-of-yet detected binary companion.

\subsection{Spectroscopy}
\label{subsec:spec}

\begin{figure*}
\centering
\includegraphics[width=0.45\linewidth]{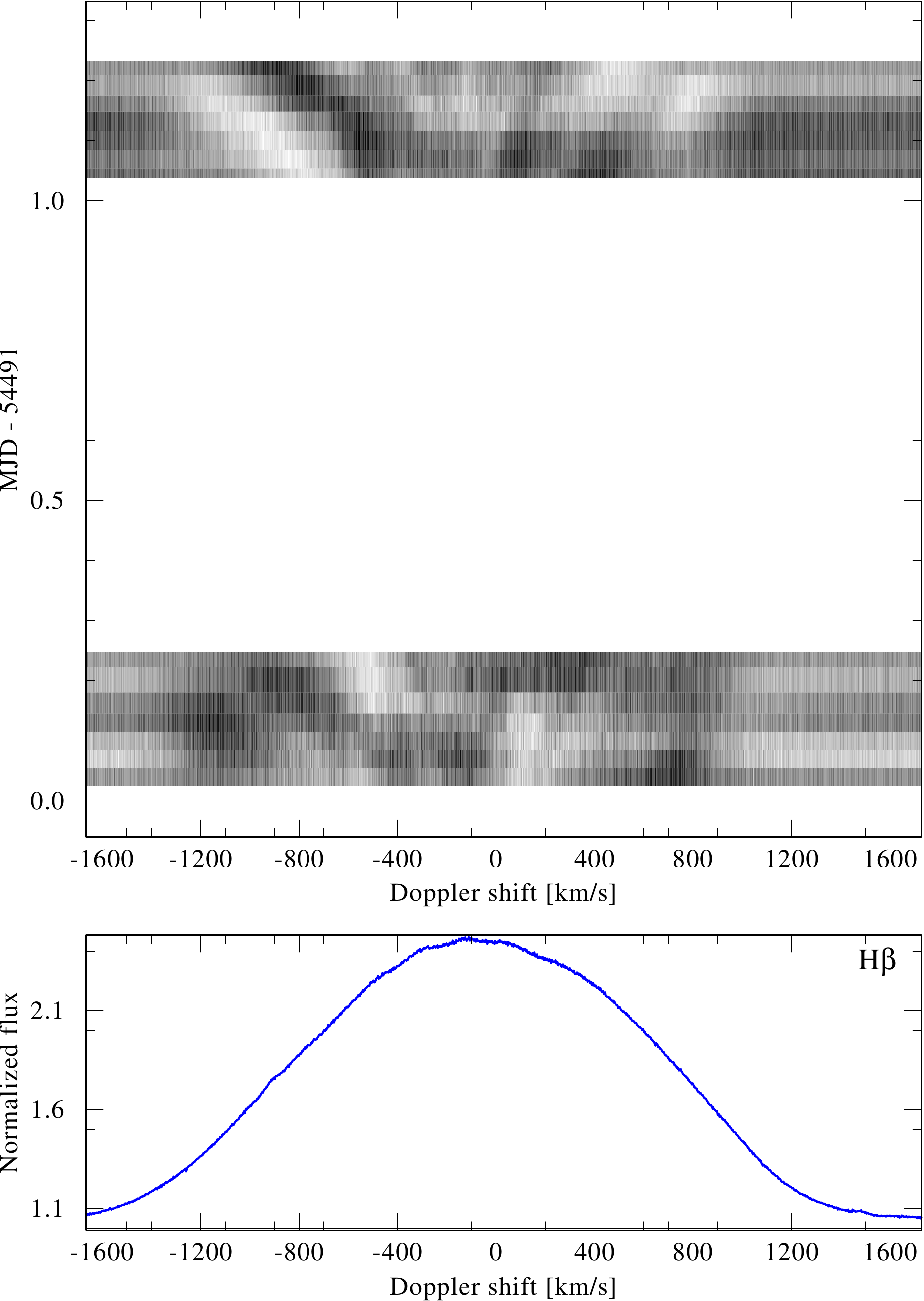}~ 
\includegraphics[width=0.45\linewidth]{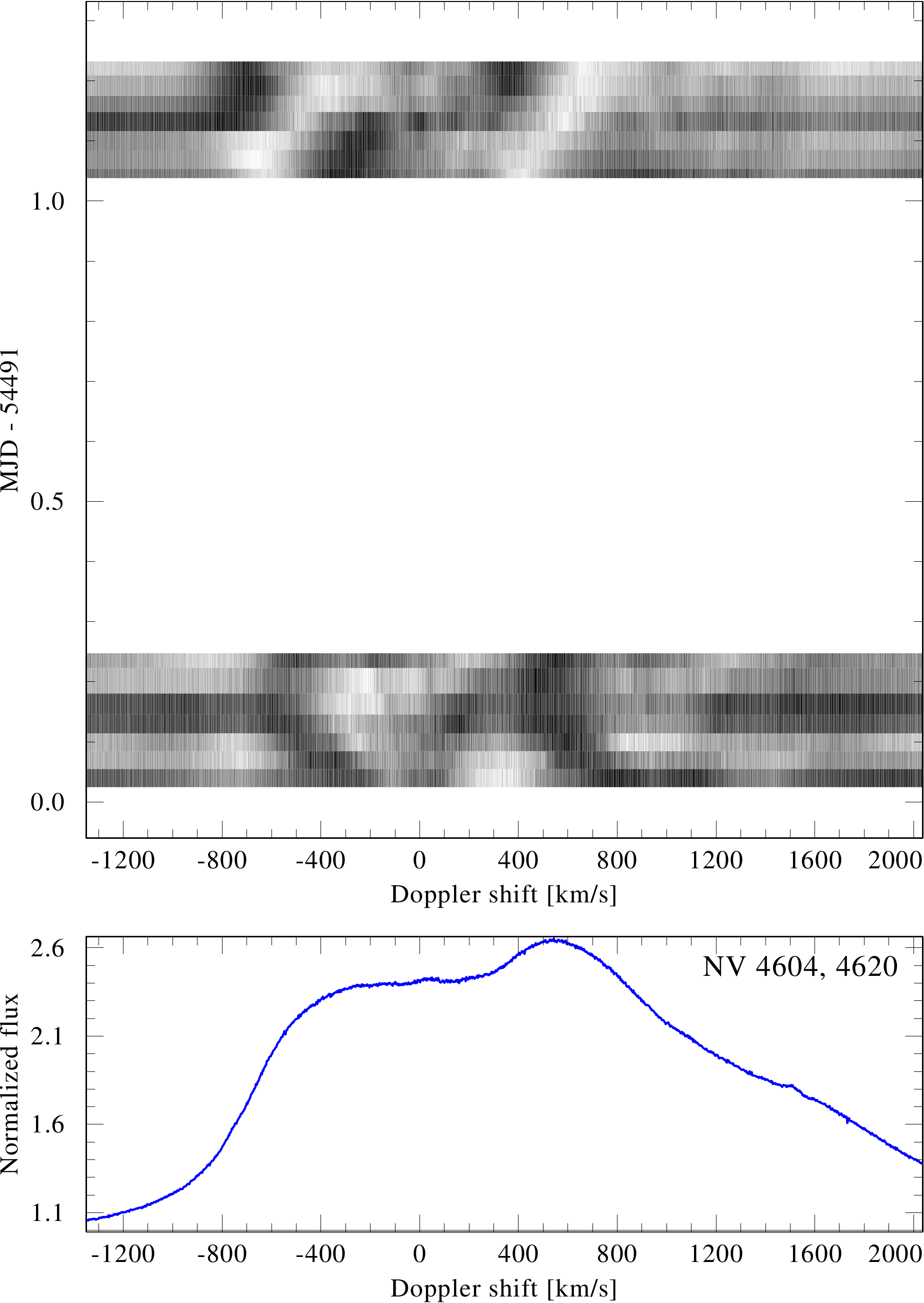} 
\caption{Residual dynamical spectra for the UVES data, zooming on the
  H\,$\beta$ (left) and the \NVblue~doublet region (right) lines. The
  panels below each dynamical spectrum show the mean spectrum
  subtracted from each observation. }
\label{fig:dynspec}
\end{figure*}

To further investigate the origin of the periods identified in the
{\itshape TESS} light curves, we show a residual dynamical spectrum of
WR\,7 using the UVES data and focusing on the \NVblue~doublet and the
H\,$\beta$ region in Fig.~\ref{fig:dynspec}. Since the UVES spectra
are of much higher SNR than the HERMES spectra and were taken on a
short cadence (see Fig.~\ref{fig:dynspec}), we refrain from combining
them with the HERMES data. The residual spectra were formed by
computing the difference between each spectrum and the mean spectrum,
plotted below the dynamical spectra for reference.

The dynamical spectra reveal structures in both these line
complexes. The H$\beta$ line shows strips of emission or absorption
propagating in Doppler space with time from the line centre. The
H$\beta$ line profile displays narrow emission peaks that travel from
near the line centre out to $\gtrsim$1000~km~s$^{-1}$ in $\sim$0.25\,d
during two consecutive nights. Such a pattern is typically associated
with clumps propagating outwards in the stellar wind
\citep{Lepine1998}. On the other hand, the N\,{\sc v} seems to suggest
a more peculiar variability. It shows preferentially negative-velocity
moving peaks during the first night and positive-velocity peaks on the
second night. The latter resembles CIR-like variability, as seen for
example in \object{WR\,1} \citep{Chene2010} or could be a
manifestation of mass transfer \citep{Packet1981} to an unseen
companion.

We attempted to phase-fold the spectra with the main periods found in the {\itshape TESS} data. However, none of these attempts resulted in coherent dynamical spectra. Notably, we cannot identify the strong 2.64~d signal obtained from the {\it TESS} data analysis in the UVES data. While we cannot determine a period from the available spectroscopic data, an inspection of the apparent CIR patterns in the N\,{\sc v}\,$\lambda \lambda 4604, 4620$ doublet (Fig.\,\ref{fig:dynspec}) suggests a possible periodicity of the order of 0.4\,d. However, continuous time coverage along a few cycles is needed to verify this.

\begin{figure}
\centering
\includegraphics[width=0.9\linewidth]{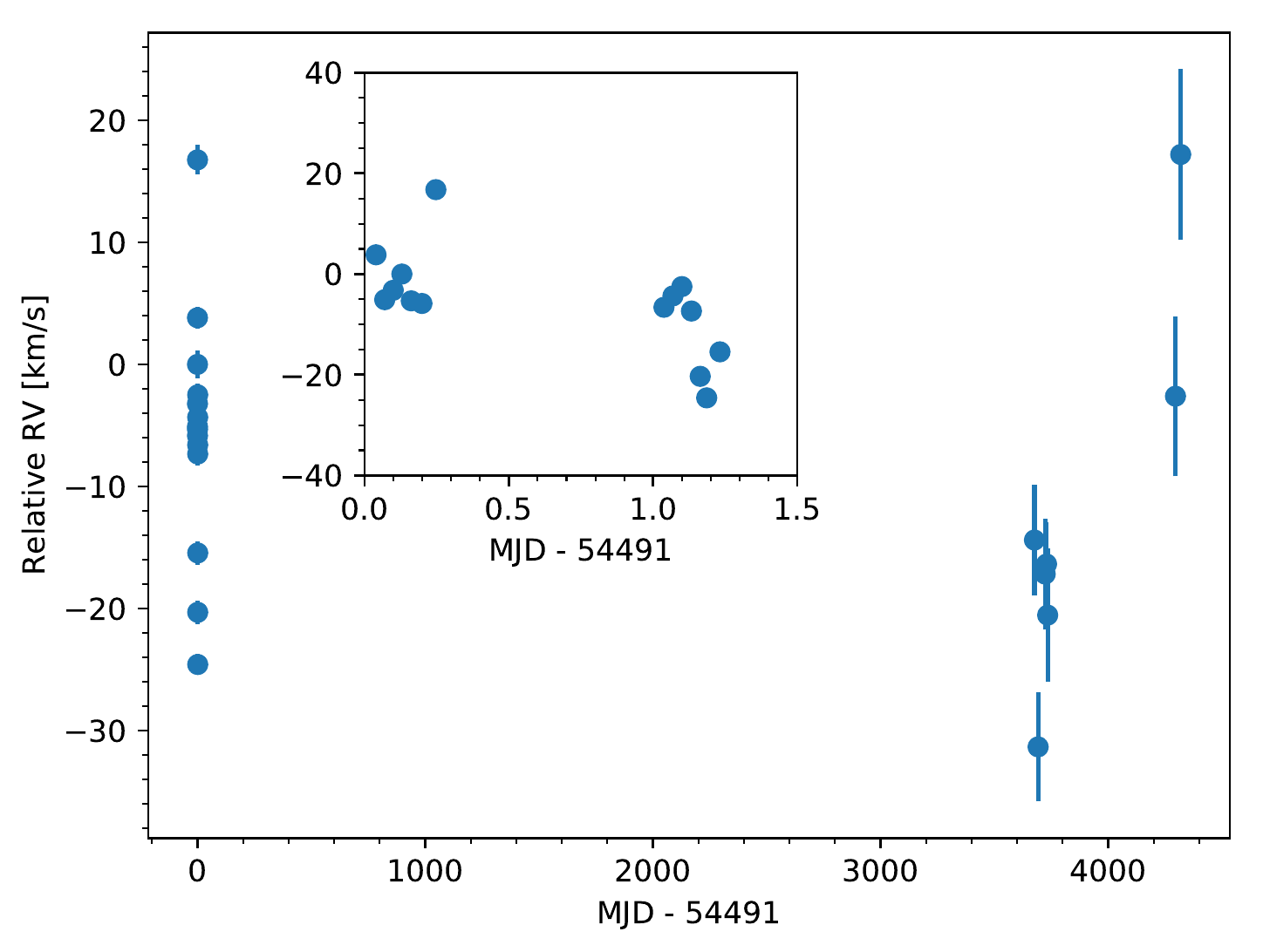} 
\caption{RVs of the \NVred~line in the available UVES and HERMES spectra. The inset shows the details of the earliest observations.}
\label{fig:RVs}
\end{figure}

To investigate the possible binary hypothesis of WR\,7 motivated by
the presence of unexpected high-frequency variability indicative of
pulsation modes in the {\it TESS} data, we measured the radial
velocities (RVs) of WR\,7 using both the UVES and HERMES data sets
(\citealt{Dsilva2020}, Dsilva et al.\ submitted to A\&A). The
measurements rely on the method of cross-correlation of the
\NVred~line with a template constructed by co-adding all observations
iteratively \citep[see][for
  details]{Zucker2003,Shenar2019,Dsilva2020}. The \NVred\, emission
line is chosen since it exhibits the least amount of variability of
all available lines. The RVs are relative, as they are measured with
respect to a co-added template with arbitrary calibration. It is
important to note that the measured RVs may be impacted by substantial
spectral variability.

The RVs of the \NVred~line, covering roughly 10 years, are shown in
Fig.~\ref{fig:RVs}. We tried to phase-fold the RVs with either $P=
2.64\,$d or $P = 0.288\,$d (the second most intense feature in the
{\it TESS} amplitude spectrum of WR\,7), but no noticeable pattern
emerged. While substantial RV variability is seen, its erratic nature
suggests that it originates in line-profile variability rather than
Doppler shifts. Moreover, no clear long-term systematic shift is
observed. Hence, while a companion cannot be ruled out from the RV
measurements, no clear evidence for a companion is seen from the RV
variability.

\begin{figure}
\centering
\includegraphics[width=\linewidth]{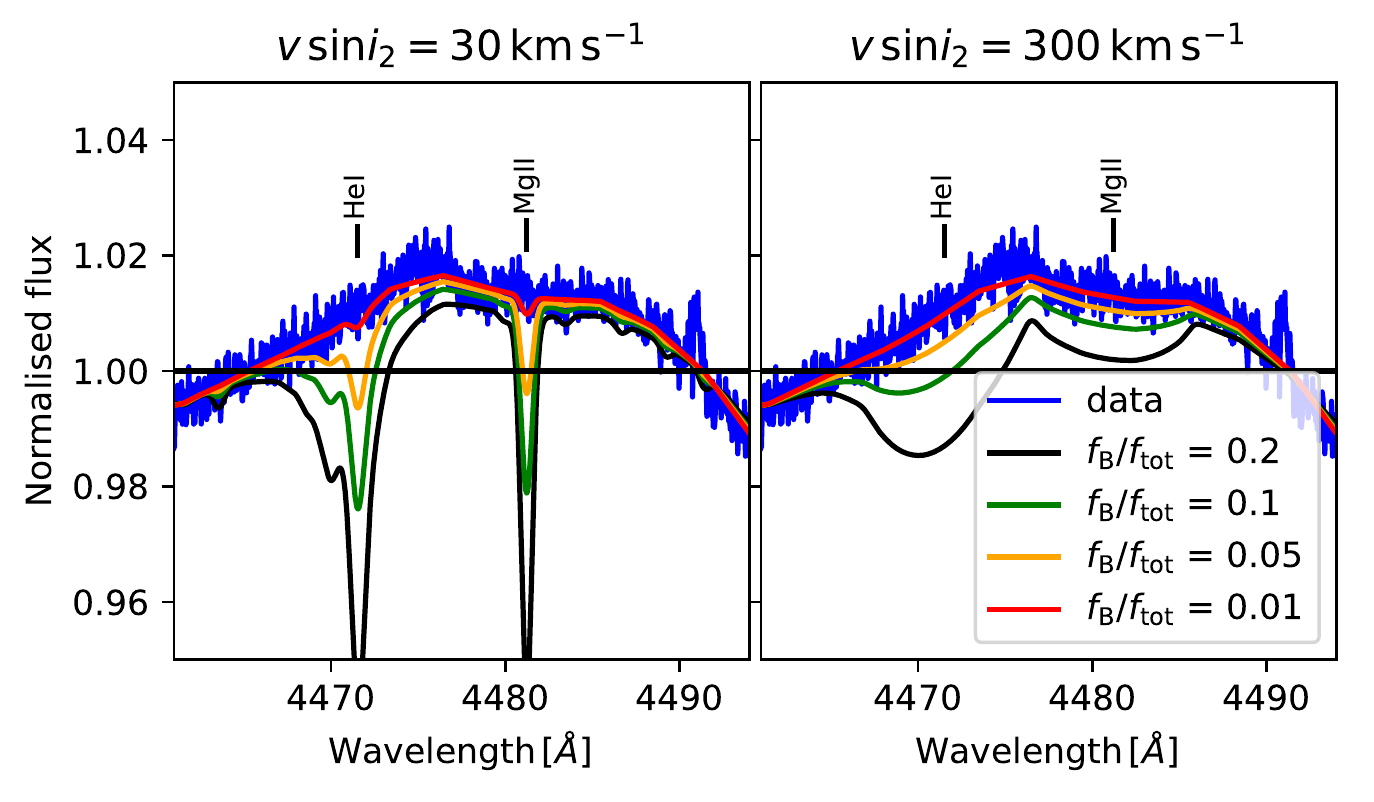} 
\caption{The two panels show the co-added UVES spectrum of WR~7,
  focusing on the He\,{\sc i}\,$\lambda 4471$ and Mg\,{\sc
    ii}$\lambda$4482 lines. Also shown are combinations of a WR
  template and a B-star template, representing a putative B-type
  companion, with varying light contributions (see legend). For the
  B-type spectrum, we use a TLUSTY model atmosphere of a late B-type
  star ($T_{\rm eff} = 15\,$kK, $\log g = 4.0\,$[cgs]). For the WR
  template, we do not use model atmospheres since they generally do
  not reproduce the underlying profile to the needed level of
  accuracy. Instead, we use the spectrum itself, binned at
  $4\,$\AA. We show this for a slow rotator ($\varv \sin i_2 =
  30\,$\kms, left panel) and a rapid rotator ($\varv \sin i_2 =
  300\,$\kms, right panel).  }
\label{fig:LightRatio}
\end{figure}

In an attempt to constrain the presence of a possible non-degenerate
stellar companion in WR\,7, we furthermore critically inspect the
co-added UVES spectrum (SNR $\approx$ 500) and compare it with
synthetic OB-type model spectra obtained from publicly available grids
calculated with the TLUSTY model atmosphere code \citep{Hubeny1995,
  Lanz2007}. Since a companion is not readily apparent in the co-added
spectrum or the RV variability, it is clear that a putative companion
cannot be of earlier spectral type (see examples in
\citealt{Shenar2019}). We therefore consider companions in the range
$T_{\rm eff} \lesssim 30\,$kK. In this regime, aside from the Balmer
lines, which would be heavily entangled with the WR emission lines,
the strongest lines are He\,{\sc i}\,$\lambda 4471$ and Mg\,{\sc
  ii}\,$\lambda 4482$, where the latter becomes stronger as the
temperature drops. As an example, Fig.~\ref{fig:LightRatio} shows the
impact a late-type B star would have on the spectrum for various light
ratios, and considering a slow and a rapid rotator. The light ratio
and the projected rotational velocity are the two limiting factors
here. As Fig.~\ref{fig:LightRatio} illustrates, we can rule out
companions contributing as little as $\approx$2--3\% if slow rotation
is assumed, which goes up to $\approx 5-10$\% if rapid rotation is
assumed.

To translate these limits into rough spectral type thresholds, we use
calibrations between spectral types and visual magnitudes by
\citet{Schmidt-Kaler1982}. Given the evident faintness of a potential
companion, we attribute the derived absolute visual magnitude of
$M_\mathrm{V}=-3.62\,$mag \citep{Hamann2019} to WR\,7. Our thresholds
therefore correspond to maximum magnitudes of $M_{\rm V, 2} \approx
0.6\,$mag for the slow rotator case and $M_{\rm V, 2} \approx
-1.1\,$mag for the rapid rotator case. These threshold magnitudes
correspond roughly to A0~V and B5~V, respectively. We note that the
very shallow and broad absorption features seen in
Fig.~\ref{fig:LightRatio}, which have a depth of the order of the SNR,
could be interpreted as photospheric absorption. However, their
presence sensitively depends on the normalisation of the data.

Thus, we can rule out companions earlier than about B5~V. A late-type
rapidly rotating B-type companion cannot be ruled out with this
analysis. A-type companions are difficult to imagine, given that the
timescale for their formation from the pre-main-sequence star
approaches or exceeds the timescale for forming a WR star. Another
alternative is that WR~7 hosts a compact object (WR+cc). As discussed
by \citet{Toala2018} for the case of WR\,124, WR+cc systems are not
necessarily X-ray bright. A NS would easily escape detection through
our RV monitoring, given the spectral variability
(cf. Fig.~\ref{fig:RVs}).

\section{Discussion}
\label{sec:discussion}

The optical data presented in this work unambiguously show that WR\,7
exhibits a range of variability time scales. The analysis of
multi-instrument observations confirms that WR\,7 has variability time
scales of the order of days and hours, which are also modulated in
scale of years. Variability in WR stars might be caused by different
factors involving companions, pulsations and density structures in
their winds. In the following we assess these possibilities as the
cause of the observed small frequencies ($<$3~d$^{-1}$) detected from
WR\,7, in particular, that corresponding to a period of 2.64~d. We
note that, as discussed by \citet{Naze2021}, the variability periods
between 3 and 10~d$^{-1}$ may be pulsations if they do not originate
from a companion.

The relatively longer period of the dominant variability
($\sim2.64$\,d) obtained from sectors 33+34 of the {\itshape TESS}
data is difficult to reconcile as pulsations \citep[see,
  e.g.,][]{Scuflaire1986,Glatzel1996M}. Theoretical work by
\citet{Grassitelli2016} who used numerical calculations of pulsating
WR stars with masses around $M_{\star}$=9--15~M$_\odot$ (covering the
mass range of WR\,7) found that their pulsation periods are of the
order of minutes. This is a much shorter-period regime than any of the
observed frequencies in the {\it TESS} light curve. Thus, we discard
pulsations from WR\,7 as the source of dominant variability
($P=2.64$~d) in sectors 33+34. In contrast, we note that the 9.8~h
period variability detected from WR\,123 \citep{Lefevre2005} has been
reproduced by radiation-hydrodynamic simulations of pulsations of
massive hydrogen-rich stellar envelopes of stars in the
22--27~M$_\odot$ range presented by \citet{Dorfi2006} \citep[see also
  the work of][]{Glatzel2008,Townsend2006}.

A large fraction of WR stars have been reported to be part of binary
systems with stellar or compact companions \footnote{See the
P.\,Crowther WR Star Catalogue \citep{Rosslowe2015}
\url{http://pacrowther.staff.shef.ac.uk/WRcat/index.php}}. \citet{vdHucht2001}
list that about 40 per cent of WR stars host binary
companions. However, we note that such percentage could be higher if
WR stars are part of long-period systems
\citep[see][]{Dsilva2020}. Thus, it would not be unusual to find that
WR\,7 is also in a binary system.

According to our spectroscopic analysis, a binary companion earlier
than B5\,V can be ruled out, but not a rapidly-rotating B-type
star. Thus, an example candidate in this restricted mass range is a Be
star. These objects are accepted to be rapidly rotating main sequence
B stars that exhibit variability due to non-radial pulsation modes
with periods between 0.01 and 10~d \citep[see, e.g., Table~1
  in][]{Rivinius2013}. Recently, \citet{Labadie2020} used {\itshape
  TESS} observations of more than 400 classical Be stars to show that
they exhibit multiperiodic variability in the frequency regime between
about $0.25 - 4$~d$^{-1}$, but can be higher in Be stars rotating near
their critical break-up velocity. Nevertheless, we note that a unique
characteristic of Be is stars is the presence of an decretion disk,
usually detected through Balmer lines in emission
\citep{Rivinius2013}, which we do not see from the WR\,7 spectroscopic
data.

There is now a large number of binary systems that have been shown to
display tidally-excited oscillations (TEOs) on a wide variety of
timescales \citep[see, e.g.,][and references
  therein]{Welsh2011,Fuller2012,Guo2017,Guo2020}. If WR\,7 possesses a
binary companion and the orbit were even slightly eccentric, TEOs may
be excited in one of the stars. Depending on their relative brightness
and the pulsation amplitudes, either one or both sets of TEOs might be
detectable.  Furthermore, given the connection that has been shown to
exist between pulsations and the formation of a structured wind
\citep{Kaufer2006, Townsend2007}, the WN TEO pulsations would
naturally give rise to the sub-features that are observed in its
emission lines. WR\,7 would not be the first WR system in which TEOs
are discovered, given recent {\itshape TESS} results on the SMC system
HD\,5980 \citep{KolaczekSzymanski2021}. But we emphasise that this
hypothesis is purely speculative given the unclear binary status of
WR\,7.

To further peer into the possible binary status of WR\,7, we
re-analyzed the available {\it XMM-Newton} EPIC data. We confirmed
that the X-ray emission from WR\,7 is soft
\citep{Zhekov2014,Toala2015} and consistent with the self-shocking
wind mechanism in WR stars \citep{Oskinova2015}. We created multi-band
light curves and found no signs of variability. Unfortunately, the
lack of variability in X-rays can not rule out the presence of compact
companions \citep{Sen2021,Toala2018}.

Other than WR\,7, there is only one case reported to exhibit
low-frequency quasi-coherent variability, namely WR\,134. A period of
2.3~d has been reported in the past decades by several authors,
suggesting that its signal is long lived
\citep{Morel1999,McCandliss1994,Naze2021}. On the other hand, the
dominant period of 2.64~d obtained from sectors 33+34 is not detected
in sector 7 observations. Even though the sector 7 data are noisy, on
average, the non-detection of the same period appears to be a robust
result.

The analysis of the {\it TESS} data of WR\,7 presented here suggests a
non-sinusoidal nature for the dominant variability, which is typically
the signature of rotational modulation. This is suggestive of the
presence of a CIR in the wind \citep[see, e.g., the case of
  WR\,110;][]{Chene_etal_2011} and, if this is the case, the real
period of the variability would be $2\times2.64$~d ($\sim5.3$~d),
given that CIR are usually present in pairs. Adopting a stellar radius
of 1.26~R$_\odot$ \citep[see][]{Hamann2019}, we estimate a rotation
velocity of 12.0~km~s$^{-1}$ for WR\,7, which is too small to be
detectable via rotational line broadening \citep[see][]{Shenar2014}.

We conclude that given the lack of strong evidence from a binary
companion in the currently available data, WR\,7 seemingly pulsates in
quasi-coherent modes in addition to variability caused by CIRs on top
of a background of stochastic low-frequency variability \citep[see,
  e.g.,][]{Bowman2019,Naze2021}. If the high-frequency variability
does originate in WR\,7 and is caused by pulsations, such a result
represents an interesting challenge to our current understanding of
the rotational, wind and pulsational properties of WR stars
\citep{Naze2021}.

\section{Conclusions}
\label{sec:conclusions}

We presented the analysis of multi-instrument optical observations of
the early WN-type star WR\,7. {\it TESS} data were used to unveil the
presence of a dominant variability with a period of
$2.6433\pm0.0005$\,d in addition to previously-reported high-frequency
features with frequencies between 3--10~d$^{-1}$, which have been
recently attributed to stellar pulsations \citep{Naze2021}.

Using high-resolution spectroscopic observations from VLT UVES we
detect small scale variations in the wind of WR\,7. The dynamical
spectra of the H$\beta$ line suggest the presence of
radially-expanding clumps in the wind of WR\,7, but that of the
N\,{\sc v} suggest negative-velocity moving peaks during the first
night and positive-velocity peaks on the second night. This could be
the manifestation of mass transfer to an unseen companion or a large
scale outflow emerging from only one hemisphere of a rotating
star. The UVES observations are further used in combination with
TLUSTY stellar atmosphere models to reject the presence of a companion
as late as A0\,V in the case of a slow rotator and a B5\,V for a rapid
rotator case. RVs measured for the WR component using the UVES data
and additional HERMES data across a 10\,yr baseline are variable, with
a peak-to-peak amplitude of $\Delta {\rm RV} \approx 50\,$\kms, but do
not give clear indication for orbital Doppler motion in WR~7.

Several mechanisms producing the dominant period in the {\it TESS}
data of sectors 33+34 of WR\,7 are explored. There are no strong
indicators of the presence of a companion, even a compact object might
be difficult to unveil \citep{Sen2021,Toala2018}. A period of 2.64~d
is difficult to reconcile with theoretical predictions from stellar
pulsation models \citep{Grassitelli2016}.

By exploring the presence of CIRs in the wind as possibility, the real
period is estimated to be $\approx 5.3$\,d. If this is the case, we
estimate a rotation velocity of 12~km~s$^{-1}$ for WR\,7, which would
be undetected given the line broadening because of the much faster
wind velocities of this WR star.

WR\,7 exhibits multi-period variability with time scales of the order
of days and hours, which are also modulated on time scales of
years. Although the presence of a stellar or compact companion cannot
be rejected, it seems that a variety of phenomena is taking place in
WR\,7. This early N-rich WR star seems to pulsate in quasi-coherent
modes in addition to variability caused by slow CIRs that might have
been overlooked in previous stellar atmosphere analysis, in addition
to stochastic low-frequency variability. WR\,7 seems to be pushing
forward our understanding of variability in evolved massive
stars. High-resolution, continuous monitoring of this star, in
addition to the most updated radiation-hydrodynamic simulations of
winds from WR stars \citep[see, e.g.,][]{Moens2022}, is needed in
order to unveil the physics behind the observed multiple variability
time scales.

\section*{Acknowledgements}

We would like to thank the anonymous referee for a detailed review on
our manuscript which improved the presentation of our analysis. JAT
acknowledges funding by DGAPA UNAM PAPIIT project IA101622 and the
Marcos Moshinsky Fundation (Mexico). DMB and TVR gratefully
acknowledge senior and junior postdoctoral fellowships from the
Research Foundation Flanders (FWO) with grant agreement numbers
1286521N and 12ZB620N, respecitively. KS and TS acknowledge funding
received from the European Research Council (ERC) under the European
Union's Horizon 2020 research and innovation programme (grant
agreement number 772225: MULTIPLES). TS further acknowledges support
from the European Union's Horizon 2020 under the Marie
Skłodowska-Curie grant agreement No 101024605. GK acknowledges support
from CONACYT grant 252499 and DGAPA UNAM PAPIIT grant IN103619.

The TESS data presented in this paper were obtained from the Mikulski
Archive for Space Telescopes (MAST) at the Space Telescope Science
Institute (STScI), which is operated by the Association of
Universities for Research in Astronomy, Inc., under NASA contract
NAS5-26555. Support to MAST for these data are provided by the NASA
Office of Space Science via grant NAG5-7584 and by other grants and
contracts. Funding for the TESS mission was provided by the NASA
Explorer Program. This research has made use of the SIMBAD database,
operated at CDS, Strasbourg, France; the SAO/NASA Astrophysics Data
System; and the VizieR catalog access tool, CDS, Strasbourg, France.

\section*{Data availability}

The data underlying this article will be shared on reasonable request to the corresponding author.


\label{lastpage}

\end{document}